\documentstyle[12pt,a4wide]{article}
\begin{document}
\input epsf
\newcommand{\be}{\begin{equation}}
\newcommand{\ee}{\end{equation}}
\newcommand{\pr}{\partial}
\newcommand{\ie}{{\it ie }}
\font\mybb=msbm10 at 11pt
\font\mybbb=msbm10 at 17pt
\def\bb#1{\hbox{\mybb#1}}
\def\bbb#1{\hbox{\mybbb#1}}
\def\bZ {\bb{Z}}
\def\bR {\bb{R}}
\def\bE {\bb{E}}
\def\bT {\bb{T}}
\def\bM {\bb{M}}
\def\bC {\bb{C}}
\def\bA {\bb{A}}
\def\bP {\bb{P}}
\def\e  {\epsilon}
\def\bbC {\bbb{C}}
\newcommand{\CP}{\bC \bP}
\newcommand{\CPP}{\bbC \bbb{P}}
\renewcommand{\theequation}{\arabic{section}.\arabic{equation}}
\newcommand{\news}{\setcounter{equation}{0}}
\newcommand{\I}{{\cal I}}
\newcommand{\HH}{\bb{H}}
\newcommand{\bx}{{\bf x}}

\title{\vskip -70pt
\begin{flushright}
\end{flushright}\vskip 50pt
{\bf \large \bf THE GEOMETRY OF POINT PARTICLES }\\[30pt]
\author{Michael Atiyah$^{\ \dagger}$ and Paul Sutcliffe$^{\ \ddagger}$
\\[10pt]
\\{\normalsize $\dagger$ {\sl Department of Mathematics and Statistics,}}
\\{\normalsize {\sl University of Edinburgh,}}
\\{\normalsize {\sl King's Buildings, Edinburgh EH9 3JZ, U.K.}}
\\{\normalsize {\sl Email : atiyah@maths.ed.ac.uk}}\\
\\{\normalsize $\ddagger$  {\sl Institute of Mathematics,}}
\\{\normalsize {\sl University of Kent at Canterbury,}}\\
{\normalsize {\sl Canterbury, CT2 7NZ, U.K.}}\\
{\normalsize{\sl Email : P.M.Sutcliffe@ukc.ac.uk}}\\}}
\date{October 2001}
\maketitle

\begin{abstract}
\noindent There is a very natural map from the configuration space of $n$ distinct
points in Euclidean 3-space into the flag manifold $U(n)/U(1)^n,$
which is compatible with the action of the symmetric group.
The map is well-defined for all configurations of points provided a certain
conjecture holds, for which we provide numerical evidence. 
We propose some additional conjectures, which imply the first, and test these
numerically.
Motivated by the above map, we define a geometrical multi-particle energy function
and compute the energy minimizing configurations for up to 32 particles.
These configurations comprise the vertices of polyhedral structures 
which are dual to those found
in a number of complicated physical theories, such as Skyrmions and Fullerenes.
Comparisons with 2-particle and 3-particle energy functions are made.
The planar restriction and the generalization to hyperbolic 3-space are also
investigated.

\end{abstract}
\newpage

\section{Introduction}\news
\label{sec-intro}

In their study of the spin-statistics theorem, Berry and Robbins \cite{BR} posed a very natural
question in classical geometry concerning the existence of a symmetric map 
between two well-known spaces. The first space, denoted by ${\cal C}_n(\bR^3)$, is the configuration
space of $n$ distinct ordered points in $\bR^3,$ and the second space is the flag manifold
$U(n)/U(1)^n$, an element of which represents $n$ orthonormal vectors in $\bC^n$, each defined
up to a phase. The Berry-Robbins problem is to construct, for each $n$, a continuous map
\be
f_n : {\cal C}_n(\bR^3)\mapsto U(n)/U(1)^n 
\label{defmap}
\ee
compatible with the action of the symmetric group $\Sigma_n$, where this acts freely
by permuting the points and the vectors respectively.

In the application of Berry and Robbins an element of ${\cal C}_n(\bR^3)$ represents the 
positions of $n$ point particles and the matrix $U(n)$ describes how a spin basis
varies as the points move in space. In this approach to the spin-statistics theorem
the Pauli sign associated with the exchange of particles arises as a geometric phase.

For the simplest case, $n=2,$ 
there is an obvious explicit map as noted by Berry and Robbins \cite{BR} but
this construction is difficult to generalize to $n>2.$ A candidate solution for all $n$ was
first presented in \cite{At1}, and is reviewed in Section \ref{sec-map}. The map is only
a candidate solution because it relies upon a certain non-degeneracy conjecture being true. 
Section \ref{sec-norm} introduces 
an appropriate determinant function (whose non-vanishing describes the non-degeneracy)
which can be
used in subsequent quantitative investigations.
In Section \ref{sec-conj} we provide numerical evidence for the validity of this conjecture and
propose and test numerically some additional conjectures, which imply the first.

Motivated by the construction of the above map, we define, 
in Section \ref{sec-min}, a geometrical multi-particle energy function
and compute the energy minimizing configurations for up to 32 particles.
Remarkably, the resulting configurations of points comprise the vertices of polyhedral structures 
which are dual to those found
in a number of complicated physical theories, including Skyrmions 
in nuclear physics and Fullerenes in carbon chemistry.
These results suggest a comparison, made in Section \ref{sec-coulomb},
 with the historic problem concerning
the minimal energy distribution of $n$ point charges on the surface of a sphere,
interacting via a 2-particle Coulomb force.
In Section \ref{sec-3pt} we propose an approximation to our multi-particle energy function in terms
of a 3-particle interaction, and 
find essentially the same minimal energy configurations.

The remaining sections concern minimal energy configurations in 
various modifications of the above picture. In Section \ref{sec-prods} we enlarge
the configuration space to consider unconstrained points in a product of spheres and show that the 
minimal energy configurations remain
unchanged. In Section \ref{sec-plane} we consider the restriction to points in the plane
and repeat our earlier comparisons. Finally, in Section \ref{sec-hyp}, we generalize
the whole situation to hyperbolic 3-space.

\section{The map}\news
\label{sec-map}
A candidate map for $f_n$ in (\ref{defmap}) was first presented in \cite{At1}, to which
we refer the reader for further details. Below we summarize the main ingredients.

First of all, any set of $n$ linearly independent vectors in $\bC^n$ can be orthogonalized,
in a way compatible with $\Sigma_n,$ so the unitarity condition in (\ref{defmap}) can be
relaxed to require a map
\be
F_n : {\cal C}_n(\bR^3)\mapsto GL(n,\bC)/(\bC^*)^n\,.
\label{defmap2}
\ee

Given 
$(\bx_1,\ldots,\bx_n)\in{\cal C}_n(\bR^3)$ then
 (\ref{defmap2}) is equivalent to defining $n$ points 
$p_i(\bx_1,\ldots,\bx_n)\in\CP^{n-1},$ for $i=1,\ldots,n,$ which are linearly
independent. We shall represent $\CP^{n-1}$ via the space of polynomials of degree
at most $n-1$ in a Riemann sphere variable $t\in\CP^1.$

The explicit map is constructed as follows. For each pair $i\ne j$ define the unit vector
\be
{\bf v}_{ij}=\frac{\bx_j-\bx_i}{|\bx_j-\bx_i|} 
\label{unitv}
\ee
giving the direction of the line joining $\bx_i$ to $\bx_j.$
Now let $t_{ij}\in\CP^1$ be the point on the Riemann sphere associated with
the unit vector ${\bf v}_{ij}$, via the identification $\CP^1\cong S^2,$ realized
as stereographic projection. Finally, set $p_i$ to be the polynomial in $t$ with
roots $t_{ij}$ ($j\ne i$), that is
\be
p_i=\prod_{j\ne i}(t-t_{ij})\,.
\label{defp}
\ee

The geometrical character of this construction means that, in addition to the
required compatibility with $\Sigma_n,$ the map is also compatible with
rotations in $\bR^3,$ where $SO(3)$ acts as the irreducible $n$-dimensional
representation on the target space. Furthermore, the map is also translation
and scale invariant; this follows trivially from (\ref{unitv}).

The reason that this map is only a candidate solution is that the following
conjecture must hold.\\

\noindent {\bf Conjecture 1}

\noindent For all $(\bx_1,\ldots,\bx_n)\in{\cal C}_n(\bR^3)$ the polynomials $p_1,\ldots,p_n$
are linearly independent.\\

For $n=2$ this conjecture is trivially true and for $n=3$ it can be proved using
simple geometry \cite{At1} or a direct algebraic computation \cite{At2}, which we
mention in the following section. 

Note that an obvious case to check is that of $n$ collinear points. Taking the line
of collinear points to be in the direction given by $t=\infty$ and  ordering
the $\bx_i$ in increasing distance along the line yields $p_i=t^{i-1},$ which are clearly independent.

For $n>3$ the conjecture remains open. In Section \ref{sec-conj} we provide numerical
evidence for this conjecture, and for some related conjectures which imply this one.
Before this, we discuss a
determinant function
 which will prove useful in making
quantitative investigations,
and which turns out to have independent interest, as we shall show.
Because of this we shall treat it in greater generality than is needed
for our immediate purposes. Readers interested in the main results
of our numerical calculations can skip the details of the next section.

\section{Determinant functions}\news
\label{sec-norm}

Linear independence can be characterized by the non-vanishing of the
appropriate determinant. Because the polynomials $p_1,\ldots,p_n$ in
conjecture 1 are only defined up to scalar factors we have to introduce
an appropriate normalization if we want a definite determinant. There are
several ways in which this can be done. One way is described in detail
in \cite{At2}: for the absolute value of the determinant one just takes
each $p_i$ to have norm 1 and then takes the volume in $\bC^n$
given by the essentially unique $SU(2)$-invariant inner product. 
The phase requires more careful treatment as explained in \cite{At2}.
There is however an alternative approach, which we shall adopt here,
that has a number of advantages. On the one hand, as already exhibited
in \cite{At2} this new definition has much better quantitative behaviour,
and this we shall be exploiting in our numerical calculations.
Another and apparently quite different advantage lies in the fact that
this new definition extends naturally to hyperbolic 3-space and hence,
on lines forecast in \cite{At2}, to Minkowski space.

We start as follows. Consider $n(n-1)$ variables $u_{ij}\in\bC^2$
($i\ne j$) $i,j=1,2,\ldots,n,$ and form the $n$ \lq polynomials\rq\
$p_1,\ldots,p_n$ given by
\be
p_i=\prod_{j\ne i}u_{ij}.
\label{m31}
\ee
This is a more abstract version of (\ref{defp}),
where $u_{ij}$ is regarded as a linear form
\be
u_{ij}=a_{ij}t_0+b_{ij}t_1
\ee
in two homogeneous coordinates $(t_0,t_1)$ related to 
the inhomogeneous coordinate $t$ of (\ref{defp})
by $t=t_0/t_1.$

If we want to avoid using coordinates, and hence
emphasize the invariance, we consider $\bC^2$ as
a vector space with a skew non-degenerate form
$(u,v).$ In particular this identifies $\bC^2$
with its dual, the space of linear forms.
Note that $\bC^2$ is the space of {\em spinors}.

In (\ref{m31}) $p_i$ is just given by the symmetrized
tensor product of $n$ copies of $\bC^2$
\be
S^n(\bC^2)\cong\bC^n.
\ee
Since $SL(2,\bC)$ acts on $\bC^2$ preserving the
skew-form it acts (irreducibly) on $\bC^n$ via
$SL(n,\bC).$

Now take the $n$ vectors $p_1,\ldots,p_n$
in $\bC^n$ and form the exterior product
\be
\omega=p_1\wedge p_2\wedge\ldots\wedge p_n
\label{m32}
\ee
which is an element of the $n$th exterior power of $\bC^n.$
Since there is a canonical isomorphism
\be
\Lambda^n(\bC^n)\cong\bC
\label{m33}
\ee
$\omega$ is essentially a complex number. More precisely
\be
\omega=\varphi\ e_1\wedge e_2\wedge\ldots\wedge e_n
\label{m34}
\ee
where $e_i$ is the monic polynomial $t^{i-1},$
or in other words $\varphi$
{\em is the determinant of the matrix of coefficients of the
polynomials} $p_1,\ldots,p_n.$
Our parameter $t$ is assumed here to come from an orthogonal,
or at least symplectic basis $(t_0,t_1)$ of $\bC^2$
(see the later discussion of symplectic representatives).

We have therefore defined a {\em complex-valued function}
$\varphi(u_{ij}).$ It has the following properties\\

\noindent(1)\
$\varphi$ is invariant under the action of $SL(2,\bC)$ on the $u_{ij}.$\\

\noindent(2)\ 
$\varphi(u_{ij}^*)=\overline{\varphi(u_{ij})},$ \ where\ 
$(a+bt)^*=(-\bar b+\bar at).$\\

\noindent(3)\ 
$\varphi(u_{\sigma(i)\sigma(j)})=\mbox{sign}(\sigma)\varphi(u_{ij}),$\
for any permutation $\sigma$ of $(1,\ldots,n).$\\

\noindent(4)\ 
$\varphi$ is a multi-linear function of the $u_{ij}.$\\

\noindent(5)\
For $n=2,$ $\varphi=(u_{12},u_{21}).$\\

\underline{Remark}:
The essential difference between this definition and the earlier one in
\cite{At2} is that here we do not use any Hermitian metric on $\bC^n,$
only the volume form. That is why we have the larger symplectic
group $SL(2,\bC)$ rather than just $SU(2).$

In terms of $\varphi$ we can proceed to define a sequence of related functions
$\varphi_k$ (for $2\le k < n$), using subsets $I$ of $(1,\ldots,n)$ of length
$|I|=k.$ For each such $I$ let $\varphi_I$ be the function $\varphi$ applied
to the variables $u_{ij}$ with $i,j\in I,$ and then put
\be
\varphi_k=\prod_I\varphi_I, \quad |I|=k.
\ee
Thus we have the sequence of functions
\be
\varphi=\varphi_n,\varphi_{n-1},\ldots,\varphi_2.
\label{m35}
\ee
Clearly from property (4) of $\varphi$ we deduce\\

\noindent(6)\ 
$\varphi_k$ is homogeneous in each $u_{ij}$ of degree
$\big({{n-2}\atop{k-2}}\big).$\\

If we take a ratio of appropriate powers of the $\varphi_k$ then we will get a
rational function of homogeneity zero in the $u_{ij}.$ This means that it is a
rational function of the corresponding points $t_{ij}\in P_1(\bC).$ In particular
we shall be interested in
\be
D(t_{ij})=\varphi_n(u_{ij})/\varphi_2(u_{ij}).
\label{m36}
\ee
Note that this has poles only where $\varphi_2(u_{ij})=0,$
ie. where $u_{ij}$ and $u_{ji}$ are proportional, or equivalently
where $t_{ij}=t_{ji}.$ From now on
{\em we restrict ourselves to the subspace of the variables where,
for all $i,j,$ $t_{ij}\ne t_{ji}.$}

A convenient way to make the definition of $D$ more explicit is to use
{\em symplectic representatives} for the $u_{ij}.$ By definition this means that
we choose each pair $u_{ij},u_{ji}$ so that (for $i<j$)
\be
(u_{ij},u_{ji})=1.
\ee
This makes $\varphi_2=1$ and so $D=\varphi$ is just the determinant of
the coefficients of the polynomials $p_1,\ldots,p_n.$

If we introduce a Hermitian metric on $\bC^2,$ with $SU(2)$ now being
the symmetry group we can introduce the anti-podal map
\be
t\mapsto t^*=-\bar t^{-1}
\ee
and we can lift this to an anti-linear map $u\mapsto u^*$ on $\bC^2.$
Explicitly, in terms of a standard basis, this is
(as in (2) above) $(a,b)\mapsto (-\bar b,\bar a).$
If we think of $\bC^2$ as the quaternions then $u^*=uj.$
Note that 
\be
(u,u^*)=|a|^2+|b|^2=|u|^2
\ee
so that if $|u|=1,$ the pair $u,u^*$ are a symplectic pair.
Such a pair we shall briefly refer to as an orthogonal pair
(since $|u|=|u^*|=1, <u,u^*>=0$).

We are now ready to return to our configurations of points
$\bx_1,\ldots,\bx_n$ in $\bR^3$ and the corresponding points $t_{ij}$
(or ${\bf v}_{ij}$) given by (\ref{unitv}), ie. by the directions
of the vectors ${\bf x}_j-{\bf x}_i.$ Our function $D(t_{ij})$ then gives
rise to a function $D({\bf x}_i)$ on ${\cal C}_n(\bR^3).$
Since our $t_{ij}$ now satisfy $t_{ij}=t_{ji}^*$ we can choose orthogonal
representatives for the $u_{ij}$ and so we get $D$ as the determinant of the
coefficients of the polynomials $p_1,\ldots,p_n.$

In \cite{At2} we defined $D$ explicitly in this way, except that we
multiplied it by a numerical coefficient $\mu(n).$ This arose from using
the invariant inner product on $\bC^n,$ but is not natural from our
present more invariant point of view. We have therefore dropped it.
Note however that the geometrical considerations in \cite{At2}
led to an upper bound for $|D|,$ which now becomes
\be
|D|\le \mu(n)^{-1}=\big\{\prod_{s=0}^{n-1}\big({n-1\atop s}\big)\big\}^{1/2}
\label{m37}
\ee
where $\big({n-1\atop s}\big)$ is the binomial coefficient.

The whole purpose of introducing our function $D$ is of course that conjecture 1
is equivalent to
\be
D(\bx_1,\ldots,\bx_n)\ne 0.
\label{m38}
\ee
Properties (1) and (2) show that, as a function ${\cal C}_n(\bR^3)\mapsto \bC$
it is covariant with respect to the full Euclidean group of $\bR^3,$
with reflections acting as complex conjugation on $\bC.$
This implies in particular that $D$ is {\em real} for any {\em planar}
configuration, which is automatic for $n=2$ ($D=1$) and $n=3.$ In general,
for $n\ge 4,$ $D$ is complex and we shall introduce its norm
\be
V=|D|
\ee
as a real-valued function on ${\cal C}_n(\bR^3)$ and refer to it briefly
as the {\em volume}.
For any collinear set we have already noted that, in a suitable orientation,
we have $p_i=t^{i-1}$ and so $V=1.$

For $n=3$ the calculation of the volume yields a nice geometrical
answer \cite{At2}. Let the triangle formed by the three points
$\bx_1,\bx_2,\bx_3$ have angles $\theta_1,\theta_2,\theta_3,$ then
\be
V=\frac{1}{2}[
\cos^2(\theta_1/2)+
\cos^2(\theta_2/2)+
\cos^2(\theta_3/2)
]\,.
\label{vtri}
\ee
This formula is obtained by explicitly computing the polynomials
$p_i$ and using some elementary geometry.

Using the fact that $\sum_{i=1}^3\theta_i=\pi$ the critical
points of $V$ are easily determined as the solutions of
\be 
\sin\theta_1=\sin\theta_2=\sin\theta_3\,.
\ee
There are two classes of solutions. The first is $\theta_1=\theta_2=0$
and $\theta_3=\pi,$ in which the triangle degenerates to three collinear
points with $V=1.$ This is the global minimum of the volume.
The second is the global maximum, given by the equilateral triangle
$\theta_1=\theta_2=\theta_3=\pi/3,$ for which $V=9/8.$
Thus, $V$ is non-zero for all configurations of three points and conjecture 1
is proved in the case $n=3.$ 

For $n>3$ conjecture 1 has yet to be proved. 
In the following section we make use of the volume function $V$ to provide numerical
evidence for this conjecture, and for some related conjectures which imply this one.

\section{Conjectures and tests}\news
\label{sec-conj}

\begin{figure}[ht]
\begin{center}
\leavevmode
\ \hskip 0cm
\ \vskip 0cm
\hbox{
\epsfxsize=7cm\epsffile{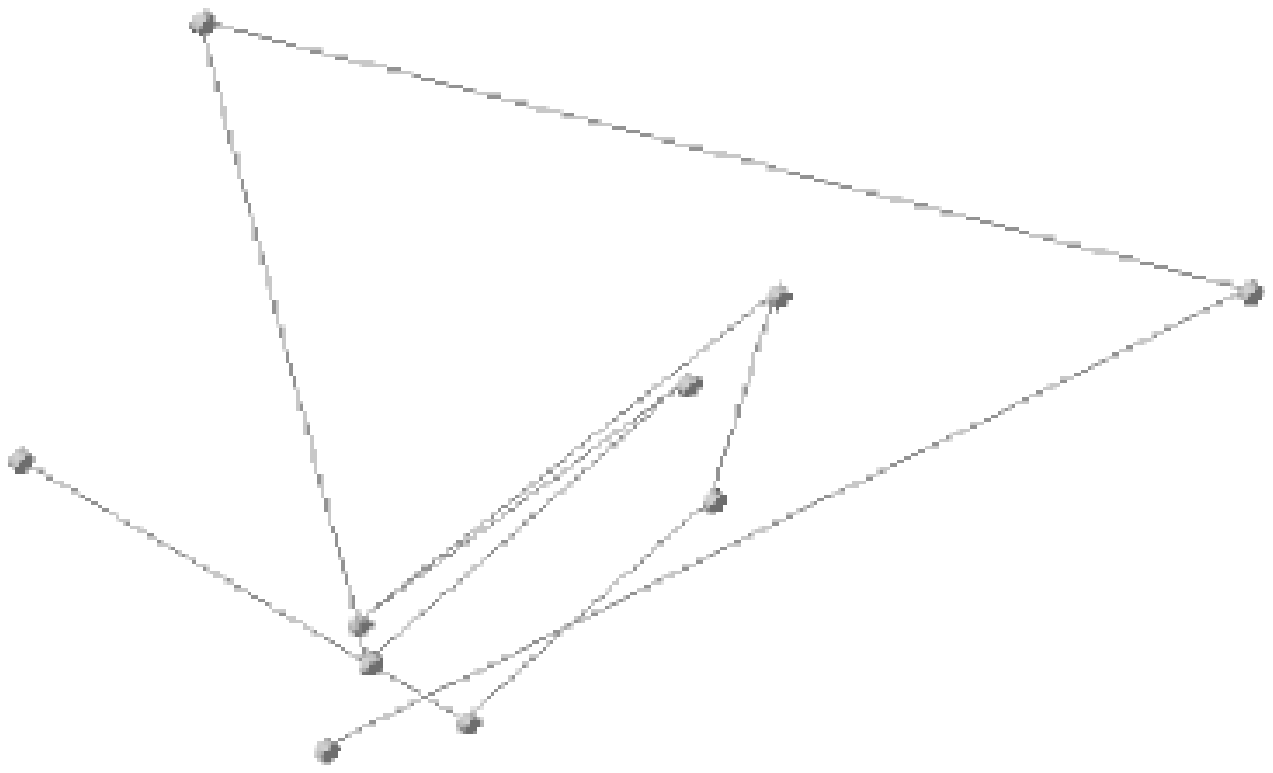}
\epsfxsize=7cm\epsffile{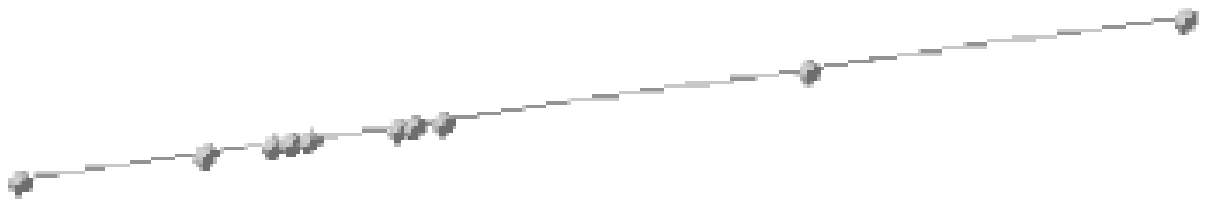}
}
\vskip -0.0cm
\caption{An initial random configuration of ten points
is shown in the left-hand plot, with lines connecting 
consecutive points to aid visualization. The right-hand
plot is the end result of applying the annealing process
to minimize the volume,
resulting in ten collinear points. }
\label{fig-conjecture}
\end{center}
\end{figure}

The calculations of the previous section prove that $V\ge 1$
for $n=2$ and $n=3.$ Furthermore, we have seen that for
$n$ collinear points, $V=1.$ This prompts us
to make the following slightly stronger conjecture.\\

\vbox{\noindent {\bf Conjecture 2}

\noindent For all $(\bx_1,\ldots,\bx_n)\in{\cal C}_n(\bR^3),$
the volume satisfies $V\ge 1.$\\}

Note that conjecture 2 implies conjecture 1, since this only
required that $V$ be non-zero. 
However the numerical evidence we provide below is all
consistent with the stronger conjecture 2.

In order to test conjecture 2 we apply a numerical
minimization algorithm known as simulated annealing 
\cite{sabook} to search the configuration space
of $n$ points for the minimum value of the function
$V$. In each case we perform several annealing
runs with the initial conditions
generated by assigning random positions to each
of the $n$ points. We have applied this procedure for all
$n\le 20$ and in each case the end result of the annealing
process is a collinear set of points with the associated
volume $V=1,$ to a high precision. 
As an example we display in fig.~\ref{fig-conjecture} 
the initial
and final configurations of points for a typical annealing
run with $n=10.$ The initial volume has the value
$V=187.07..,$ and at the end of the annealing process
this has reduced to $V=1.00000..,$ with the associated
points being collinear.
We believe that this evidence for the
conjecture is quite convincing.

One obvious line of attack for proving conjecture 2 would be to
attempt some form of proof by induction, since we already know that
this conjecture holds for $n=2$ and $n=3.$ As a move in this direction
we propose a further conjecture.

To state this recall the sequence of functions $\varphi_k$ we introduced
in (\ref{m35}). We noted that the ratios of appropriate powers of the
$\varphi_k$ would be functions of the $t_{ij}$ (and hence functions on
${\cal C}_n(\bR^3)$) and we defined $D$ by (\ref{m35}) as the ratio of
$\varphi_n$ to $\varphi_2,$ and equal to $\varphi_n$ in the symplectic
normalization which makes $\varphi_2=1.$ Since $\varphi_{n-1}$ has
homogeneity $(n-2)$ (property (6)) we can also consider
\be
\varphi^{n-2}/\varphi_{n-1}.
\label{m41}
\ee
This can also be rewritten as
\be
D^{n-2}/\prod_{i=1}^nD_i
\label{m42}
\ee
where $D_i$ is the function $D$ evaluated on the configuration obtained
by omitting ${\bf x}_i.$ In terms of absolute values this gives the function
(with $V_i=|D_i|$)
\be 
\chi=\frac{V^{n-2}}{\prod_{i=1}^n V_i}.
\label{m43}
\ee
Our third conjecture can now be stated.\\

\noindent{\bf Conjecture 3}

Let $V$ denote the volume for $n$ points, 
$(\bx_1,\ldots,\bx_n)\in{\cal C}_n(\bR^3),$
 and let $V_i$ denote the volume for the $n-1$ points
$(\bx_1,\ldots,\bx_{i-1},\bx_{i+1},\ldots,\bx_n)\in{\cal C}_{n-1}(\bR^3),$
 obtained by deleting the point $\bx_i.$ Then, for
 all $(\bx_1,\ldots,\bx_n)\in{\cal C}_n(\bR^3),$ 
\be V^{n-2}\ge \prod_{i=1}^n V_i.
\label{conj3}\ee

\ \\

Conjecture 3, applied inductively, eventually shows that a power of $V$
is bounded below by the product of volumes over all pairs $(\bx_i,\bx_j)$
and this is just 1. Thus conjecture 3 implies conjecture 2. In terms
of the sequence $\varphi_k$  conjecture 3 asserts that,
for the products of the appropriate powers, we get a descending sequence of
values, beginning with $V$ and ending with 1.

We have obtained similar numerical evidence for conjecture 3 as
we did for conjecture 2, by applying our simulated annealing
algorithm to the function 
$\chi$ defined by (\ref{m43}).
Again this results in a set of collinear points, for which
$\chi=1$ to a high precision and the inequality (\ref{conj3})
becomes an equality.

\section{Minimal energy configurations}\news
\label{sec-min}

In this section we investigate the configurations of points
for which the volume $V$ is maximal. For this
purpose it is convenient to introduce the energy function
\be
E=-\log V 
\label{energy}
\ee
so that the critical points we seek are the minimizers of this
energy. This is quite a natural definition of the
energy in that, if we consider two well-separated clusters, then
the volume factorizes and so the total energy is the sum of
the energies of the two non-interacting clusters. 
In fact the deviation from the sum is of order $1/r,$ where
$r$ is the ratio of the separation to the cluster scale, 
but there is also an angular dependence.

As Gary Gibbons has pointed out to us, the definition
(\ref{energy}) involving the logarithm of a volume 
suggests that it may be interesting to explore the interpretation 
of this quantity in terms of entropy.

We see that in our geometrical investigations of 
point configurations or \lq particles\rq\ 
we have been led to an interesting multi-particle
energy function which places a penalty on compressing the 
associated volume form. 
We shall now investigate the 
minimal energy arrangements of points.
Note that the upper bound (\ref{m37}) for $|D|$  
gives a lower bound for the energy $E.$
  
Note that conjecture 2 is equivalent to the statement
that the energy is non-positive, $E\le 0,$ whereas 
conjecture 1  merely implies that the energy is finite.

For two points, the energy is independent of the positions
of the points, it being identically zero, 
so there is no 2-particle interaction energy.

For three points, we have already seen that the minimal
energy configuration is an equilateral triangle (with
arbitrary scale, location and orientation) and the
energy is $E=-\log(9/8)=-0.11778..$.

We now apply our simulated annealing algorithm to the
energy function (\ref{energy}) to determine the structures
for larger $n$ which generalize the equilateral triangle 
at $n=3.$

\begin{figure}[ht]
\begin{center}
\leavevmode
\ \hskip -2cm
\epsfxsize=15cm\epsffile{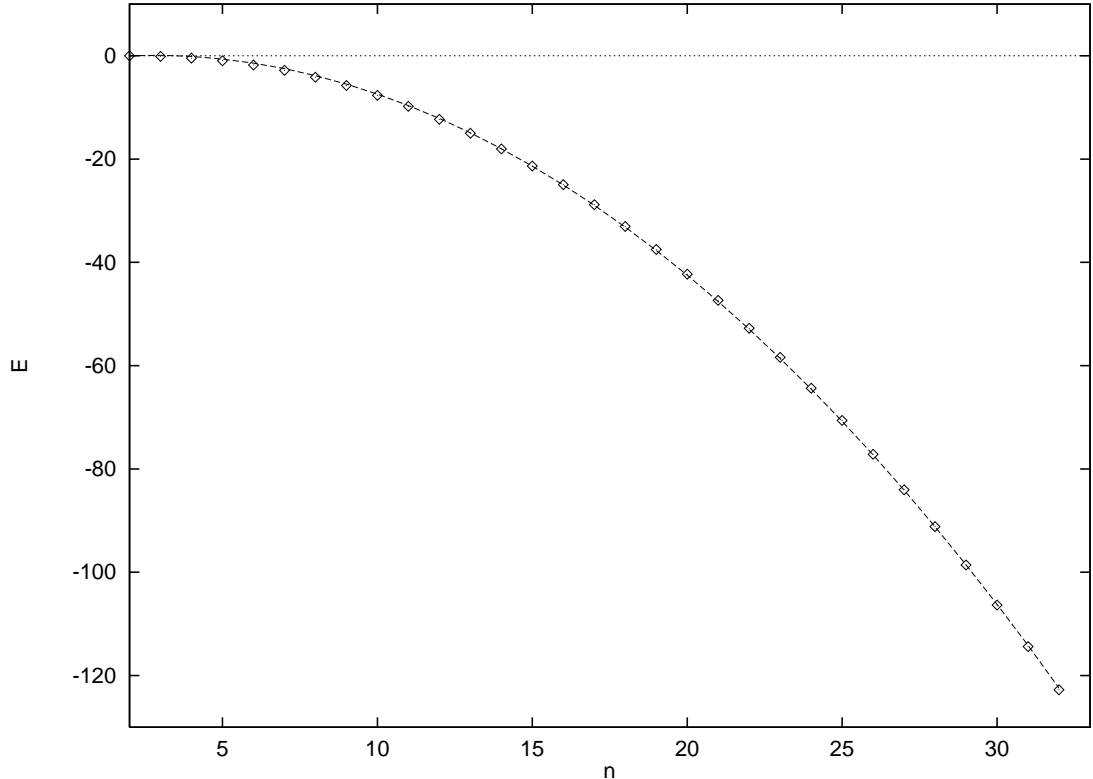}
\vskip -0.5cm
\caption{
The energy $E$ (diamonds) of the minimizing configuration
for $n$ points with $2\le n\le 32.$ The dashed curve 
represents the quadratic fit described in the text.
}
\label{fig-energy}
\end{center}
\end{figure}

\noindent In fig.~\ref{fig-energy} we plot (diamonds) the minimum value of the energy
for $2\le n\le 32.$ The precise values are listed in the second column of 
table~1. 
The dashed curve in fig.~\ref{fig-energy} is the result of a least squares
fit to the data using the quadratic approximation
\be
E(n)=-an^2+bn+4a-2b 
\label{fit}
\ee
where the constant term has been chosen so that $E(2)=0,$ in agreement
with the result that the energy of any two points is zero.
The values obtained by fitting to the data are $a=0.143$ and $b=0.792.$
As can be seen from the plot, this approximation is fairly accurate,
and it would be nice to have some understanding of this quadratic growth.

Related to this last issue, note that conjecture 3 corresponds to
an upper bound for the $n$-particle energy in terms of the 
$(n-1)$-particle energy. Explicitly,
\be
E\le \frac{1}{n-2}\sum_{i=1}^n E_i 
\label{enrec}
\ee
where $E_i$ denotes the energy when the point $\bx_i$ is removed.
If we denote by $B_{n-1}$ an upper bound for the $(n-1)$-particle energy
then (\ref{enrec}) implies that we may take $B_n=\frac{n}{n-2}B_{n-1}$ as
an upper bound for the $n$-particle energy.
Iterating this relation up from the 2-particle energy just reproduces
the result that the energy is non-positive but iterating up from
the 3-particle energy yields
\be
B_n=\frac{-n(n-1)}{6}\log(9/8) 
\label{bound}
\ee
for $n\ge 3.$
Although the numerical values in the upper bound (\ref{bound})
are poor in comparison with the fit (\ref{fit}), the quadratic
growth, damped by a linear factor, is reproduced.

\begin{figure}[ht]
\begin{center}
\leavevmode
\ \hskip -2cm
\epsfxsize=12cm\epsffile{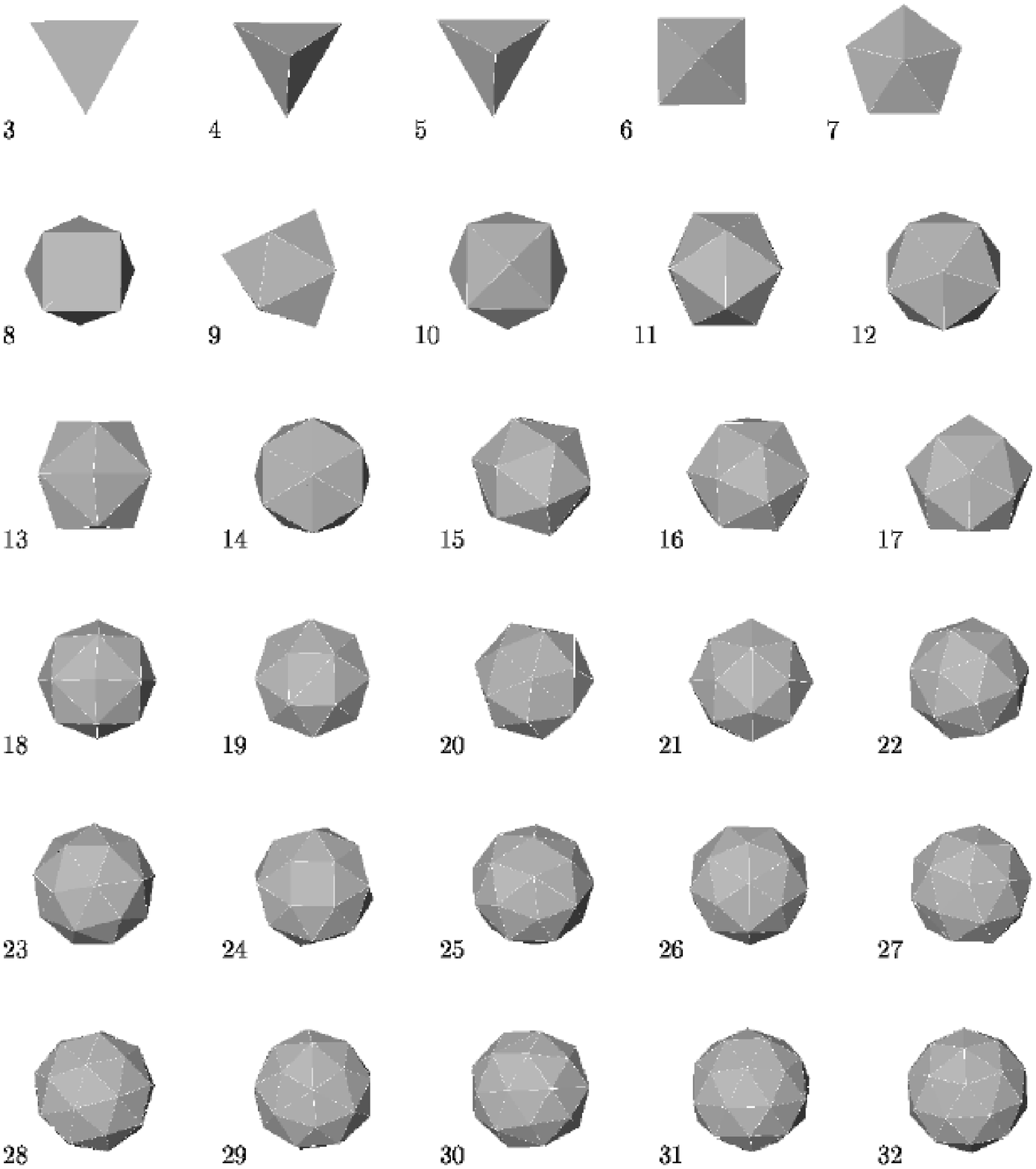}
\vskip -0.5cm
\caption{
View down the main symmetry axis of the polyhedra associated with
the energy minimizing configurations for $3\le n\le 32.$
}
\label{fig-top}
\end{center}
\end{figure}

\begin{figure}[ht]
\begin{center}
\leavevmode
\ \hskip -2cm
\epsfxsize=12cm\epsffile{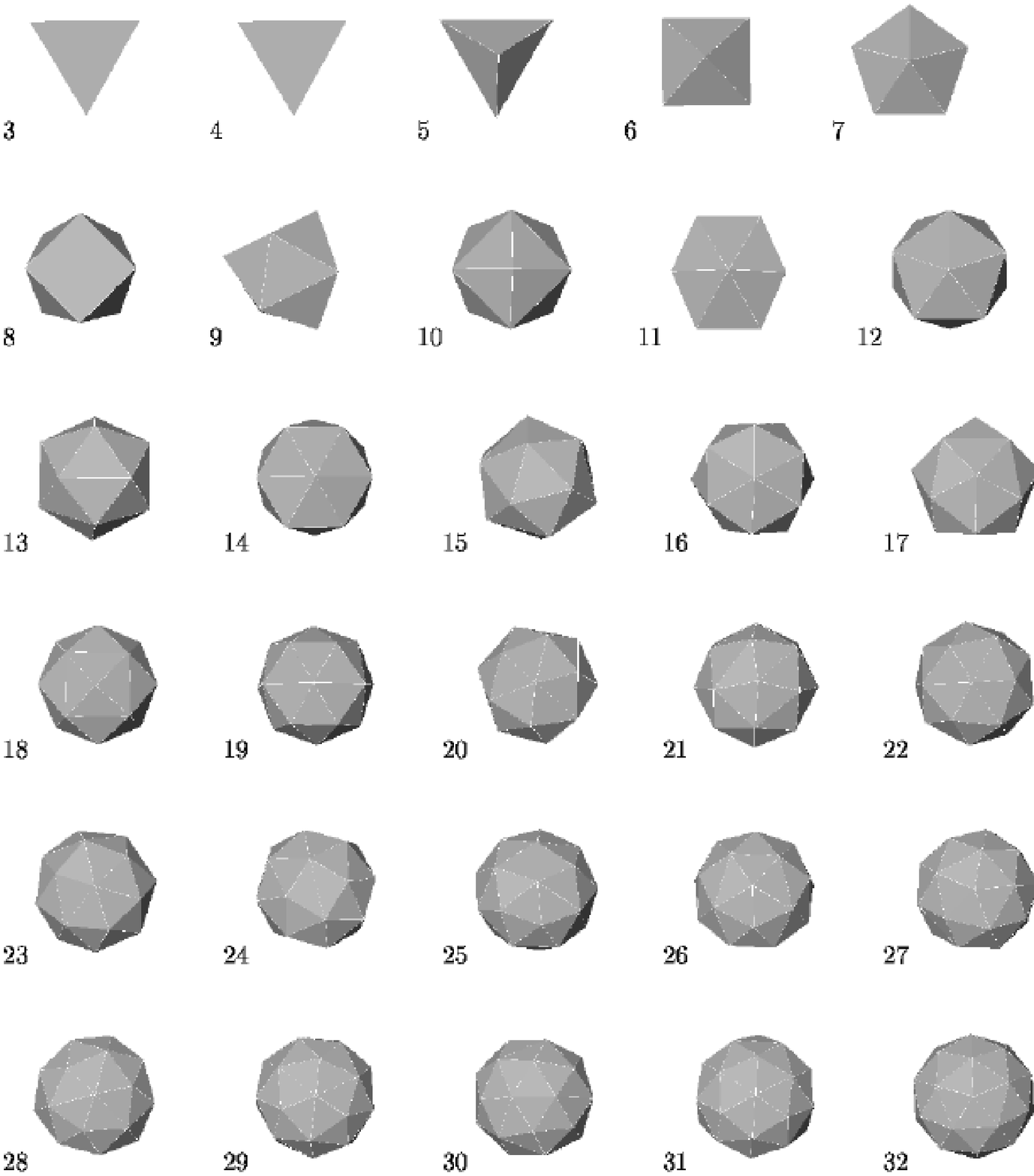}
\vskip -0.5cm
\caption{
View up the main symmetry axis of the polyhedra associated with
the energy minimizing configurations for $3\le n\le 32.$
}
\label{fig-bot}
\end{center}
\end{figure}

In the third column of table~1 we present the symmetry group of the
energy minimizing configuration and in fig.~\ref{fig-top} we display
polyhedra whose vertices consist of the $n$ points of this
configuration. The views in fig.~\ref{fig-top} are down the main symmetry
axis and the corresponding views up the main symmetry axis are presented
in fig.~\ref{fig-bot}

\begin{table}
\centering
\begin{tabular}{|r|r|c|c|r|r|c|}
\hline
$n$ & $E$\ \ \ \ \  & $G$ & $\delta$ & $E_\Delta$\ \ \  &
 $\widetilde E$\ \ \ \ \  & $\widetilde \delta$\\
\hline
   2&          0.0000&  $D_{\infty h}$&   1.00 & -\ \ \ \  & - \ \ \ \ & -\\
   3&         -0.1178&  $D_{3h}$&         1.00 & -0.1178 & -0.1178 & 1.00  \\
   4&         -0.4463&  $T_d$&            1.00 & -0.1178 & -0.4463& 1.00  \\
   5&         -0.9718&  $D_{3h}$&         0.97 & -0.1086 & -0.9708& 0.98  \\
   6&         -1.7994&  $O_h$&            1.00 & -0.1062 & -1.7994& 1.00  \\
   7&         -2.8262&  $D_{5h}$&         0.97 & -0.1018 & -2.8226& 0.98  \\
   8&         -4.1632&  $D_{4d}$&         1.00 & -0.0997 & -4.1627& 1.00  \\
   9&         -5.7746&  $D_{3h}$&         0.99 & -0.0978 & -5.7743& 0.99  \\
  10&         -7.6597&  $D_{4d}$&         0.99 & -0.0962 & -7.6591& 1.00  \\
  11&         -9.8001&  $C_{2v}$&         0.98 & -0.0947 & -9.7985& 0.99  \\
  12&        -12.3165&  $Y_h$&            1.00 & -0.0939 & -12.3165& 1.00  \\
  13&        -15.0021&  $C_{2v}$&         0.99 & -0.0927 & -15.0010& 0.99  \\
  14&        -18.0354&  $D_{6d}$&         0.99 & -0.0919 & -18.0334& 0.99  \\
  15&        -21.3443&  $D_3$&            0.99 & -0.0912 & -21.3427& 1.00  \\
  16&        -24.9525&  $T$&              1.00 & -0.0906 & -24.9498& 1.00  \\
  17&        -28.8498&  $D_{5h}$&         1.00 & -0.0900 & -28.8498& 1.00  \\
  18&        -33.0439&  $D_{4d}$&         1.00 & -0.0895 & -33.0438& 1.00  \\
  19&        -37.5018&  $C_{2v}$&         1.00 & -0.0890 & -37.4987& 1.00  \\
  20&        -42.3013&  $D_{3h}$&         1.00 & -0.0886 & -42.3005& 1.00  \\
  21&        -47.3714&  $C_{2v} $&        1.00 & -0.0882 & -47.3714& 1.00  \\
  22&        -52.7464&  $T_d$&            1.00 & -0.0879 & -52.7461& 1.00  \\
  23&        -58.3834&  $D_3$&            1.00 & -0.0876 & -58.3820& 1.00  \\
  24&        -64.3697&  $O$&              1.00 & -0.0873 & -64.3694& 1.00  \\
  25&        -70.6018&  $C_{1h}$&         1.00 & -0.0870 & -70.6010& 1.00  \\
  26&        -77.1541&  $C_2$&            1.00 & -0.0867 & -77.1541& 1.00  \\
  27&        -84.0314&  $D_{5h}$&         1.00 & -0.0865 & -84.0314& 1.00  \\
  28&        -91.1685&  $T$&              1.00 & -0.0863 & -91.1685& 1.00  \\
  29&        -98.5921&  $D_3$&            1.00 & -0.0861 & -98.5915& 1.00  \\
  30&       -106.3488&  $D_2$&            1.00 & -0.0859 & -106.3488& 1.00  \\
  31&       -114.3918&  $C_{3v}$&         1.00 & -0.0857 & -114.3917& 1.00  \\
  32&       -122.7781&  $Y_h$&            1.00 & -0.0855 & -122.7781& 1.00  \\
\hline
\end{tabular}\label{table}
\caption{For $2\le n\le 32$ we list the energy, $E,$ of the minimizing
configuration, its symmetry group, $G,$ and the deformation, $\delta,$ from
a spherical arrangement. $E_\Delta$ is the minimized average 3-particle energy 
(described later),
and $\widetilde E,\widetilde \delta$ are the values of $E,\delta$ computed from
the 3-particle energy minimizing configurations.}
\end{table}

 In case the reader is not familiar
with the notation used for point group symmetries
 we briefly recount the main details here.
The Platonic groups are the rotational symmetries of the
tetrahedron ($T$), the octahedron/cube ($O$) and the
icosahedron/dodecahedron ($Y$).
The dihedral group $D_n$ is obtained from the cyclic group of
order $n$, $C_n$, by the addition of a $C_2$ axis which is orthogonal
to the main $C_n$ symmetry axis. 
The group $D_n$ can  be extended by the addition of a reflection
symmetry in two ways: by including a reflection in the plane
perpendicular to the main $C_n$ axis, which produces the group $D_{nh}$
or, alternatively, a
reflection symmetry may be imposed in a plane which contains the main
symmetry axis and bisects the $C_2$ axes, which results
in the group $D_{nd}.$ In the same way as for the dihedral groups
the Platonic groups may also be enhanced by reflection symmetries,
again denoted by the subscripts $h,d.$ The addition of a subscript
$h$ to a cyclic group denotes a horizontal reflection symmetry,
but a vertical reflection plane is denoted by a subscript $v.$

We find that in all the above minimizing configurations the points lie on,
or very close to, the surface of a sphere. To measure the deformation from
a spherical arrangement we compute the following quantity $\delta.$
First of all, we compute the centre of mass of the configuration
${\bf X}=\frac{1}{n}\sum_{i=1}^n {\bf x}_i$ and use the translational invariance
of the problem to position this at the origin. Next we use the scale
invariance to rescale the position vectors of all the points 
(by the same amount) so that the point (or points) which is furthest
from the origin lies on the unit sphere. $\delta$ is then defined as
the distance from the origin of the point which is closest to the origin.
Clearly from this definition $\delta\le 1,$ with equality if and only
if all $n$ points lie on the unit sphere (after we have made the above
transformations). The computed value of $\delta$ is given 
 in the fourth column of table~1. In each case $\delta$ is
very close to unity, the greatest deviation occuring for $n=5$ and $n=7$
where $\delta=0.97,$ which is still quite close to a spherical arrangement.
In order to accurately compute $\delta$ the minimum energy needs to be calculated
to a high precision, but we believe that the results for $\delta$
 quoted to two decimal places 
in the fourth column of table~1 are accurate to this level. 
As an example, the $n=5$ polyhedron is a trigonal bipyramid composed of
a point at each of the north and south poles of the unit sphere and an
equilateral triangle in the equatorial plane but on a circle of radius
$0.97.$ If we rescale the equilateral triangle so that all five points
lie on the unit sphere then the energy increases from $E=-0.9718$ to
$E=-0.9714,$ indicating that the non-spherical arrangement is the correct
minimum of the energy. For $n=7$ the situation is similar, with the
polyhedron being a pentagonal bipyramid, but this time it is the two
points at the poles which are inside the unit sphere when 
the pentagon is scaled so that its vertices lie on the equator of the unit
sphere. Again rescaling to a spherical arrangement slightly 
increases the energy from $E=-2.8262$ to $E=-2.8255.$

It is perhaps useful to briefly describe the salient features of the
polyhedra we have found. For $n=4,6,12$ the polyhedra are the Platonic solids,
namely the tetrahedron, octahedron and icosahedron, respectively.
For $n=5$ and $n=7,$ we have already mentioned that the trigonal and
pentagonal bipyramids are formed. Note that the $n=6$ case also fits in
the middle of this pair, since the octahedron may be thought of as a special
case of the square bipyramid. $n=8$ is the first example in which some of the
faces are not triangular, it being a square anti-prism, obtained from a cube
by rotating the top face by $45^\circ$ relative to the bottom face. 
This example demonstrates a general feature that the most symmetric 
configurations are not automatically those of lowest energy. Nine
points lie on the vertices of three parallel equilateral triangles, with the
middle triangle rotated by $60^\circ$ relative to the other two.
The $n=10$ polyhedron can be obtained from the $n=8$ one by replacing each
square by a hat made from four triangles with a tetravalent vertex.
The first polyhedron containing a hexamer (a vertex with six nearest neighbours)
occurs at $n=11.$ There are two tetravalent vertices and the remaining eight
are pentamers (vertices with five nearest neighbours). The existence of the
single hexamer clearly forbids the solution from having much symmetry. 
Another general pattern is that if the number of points is one more or less
than an exceptionally symmetric configuration (recall that the polyhedron
for $n=12$ is the icosahedron) then the minimizing configuration tends to
have rather low symmetry. 

For $n\ge 12$ most of the polyhedra consist of $2(n-2)$ triangular faces
with $12$ pentamers and $n-12$ hexamers. Particularly symmetric examples
are the icosahedron at $n=12$ and the dual of the truncated icosahedron
at $n=32.$ Within the range we have studied,
$12\le n\le 32,$ there are six exceptions to the above rule, 
at $n=13,18,19,21,24,25.$ As can be seen from figs.~\ref{fig-top},\ref{fig-bot},
the polyhedra for $n=13,18,21$ contain tetravalent vertices (one each
for $n=13,21$ and two in the case of $n=18$) and for $n=19,21,24,25$ there
are rectangular faces (one each for $n=19,21,25$ and six squares for
$n=24$, which is a slightly deformed snub cube).

The Platonic solids with trivalent vertices (the cube at $n=8$ and
the dodecahedron at $n=20$) are clearly not favoured by the desire for
triangular faces and the formation of pentamers and hexamers, so it is
not surprising that these highly symmetric configurations do not arise.

Polyhedra (or their duals) with $n$ vertices and $2(n-2)$ triangular faces
forming $12$ pentamers and $n-12$ hexamers appear to be generic configurations
of points which arise in a number of diverse applications. Examples
include carbon chemistry, where the dual polyhedra appear with vertices
representing the positions of the carbon atoms in closed cages known
as Fullerenes \cite{KHOC}, in biology, where they arise in the structure of 
spherical virus shells \cite{BSTK}, and 
in Skyrmions, which are topological solitons which model nuclei.
In this last application the vertices
of the dual polyhedra represent points at which the baryon density is maximal.
The relation between the number of points $n$ and the baryon number $B$
is $n=2B-2.$ Comparing the results in this paper with those
in \cite{BS} we find that for $2\le B\le 17,$ which is the range
of baryon numbers for which both sets of results are known, there is an exact match
between the symmetries and combinatorial types of the Skyrmion polyhedra and
the duals of the polyhedra presented here, for all but three cases
($B=5,9,10$), and in some of these cases there is a match with known
low energy local minima Skyrmions, whose energies are extremely close to 
those of the global minima. 

Rather surprisingly, it appears that the arrangement of point charges
on a sphere is closely related to the configurations of points we have
generated by our purely geometrical construction. The details of this 
comparison are addressed in the following section.

\section{Comparison with charges on a sphere}\news
\label{sec-coulomb}

The problem, generally attributed to J.J. Thomson \cite{JJT},
 is to find the configuration
of $n$ point charges on the surface of a sphere such that the total
Coulombic energy
\be
E_1=\sum_{i>j}^n \frac{1}{|{\bf x}_i-{\bf x}_j|} 
\label{coul}
\ee
is minimal. 
This is a notoriously difficult problem, but the use of modern computers
has provided numerical results for $2\le n\le 112,$ see for example 
\cite{EH} and references therein. Remarkably we find that for all
$2\le n\le 32$ the symmetries of the configurations we have found are
identical to those of the configurations of points on a sphere which
minimize the energy (\ref{coul}) 
(compare table~1 with table~1 of \cite{JRE}).
Furthermore, the combinatorial types of the polyhedra also match.
Note that it is not true that all the configurations are identical in
each case, since our points are not constrained to lie on the
surface of a sphere, and in some cases it appears that they definitely
do not. However, as we have seen, in all cases the points lie very close
to the surface of a sphere and so it is possible to consider projecting them
onto the sphere. Since there is always an ambiguity in performing such a
projection we have instead adopted an approach which is mathematically 
better defined, namely, we constrain the points to lie
on the surface of the unit sphere throughout the minimization process.
Practically the two approaches agree, since the resulting
minimal energy configurations are identical to those obtained by  
projection of the previous configurations, to within the accuracy that we work
 (associated with the values listed in table~1). 

These configurations of points on a sphere agree with those 
that minimize the Coulomb energy, to within the accuracy that we work.
For example, we have taken each of our minimal energy configurations
and computed their Coulomb energies (\ref{coul}) and compared these
with the table presented in \cite{EH}, where we find an agreement
to at least five significant figures in each case.
 Also, comparing
figs.~\ref{fig-top} and \ref{fig-bot} with the corresponding figures
in \cite{JRE} it can be verified that the combinatorial types are
identical. This correspondence is intriguing, given our purely
geometric construction, which is scale invariant and certainly has no
explicit 2-particle interaction, but yet appears to generate an interaction which
confines the particles close to a sphere and has the same affect as the Coulomb force.
It is true that for a small number of points $n\le 6,$ the minimal arrangement  
is essentially independent of the force law and is determined
by symmetry alone, but for $n\ge 7$ the
precise arrangement and symmetry of the minimizing configuration
is sensitive to the energy formula used. For example, if the Coulomb 
interaction (\ref{coul}) is replaced by a more general power law
\be
E_p=\sum_{i>j}^n \frac{1}{|{\bf x}_i-{\bf x}_j|^p} 
\label{gen}
\ee
then the symmetry and structure of the minimal energy configuration
can be studied as a function of $p$ (and $n$) with highly non-trivial
results \cite{MKS}.
We shall return to this point later in Section \ref{sec-plane}.
 As an illustration, the Tammes problem (to determine
the configuration of $n$ points on a sphere so that the minimum distance
between the points is maximized) emerges from the potential
(\ref{gen}) in the limit $p\rightarrow\infty$ and numerical results show
that for $n>6$ the only configuration which is a common solution
of the Coulomb and Tammes problem is the icosahedral arrangement for $n=12$
\cite{EH}. This makes our observed matching with the Coulomb problem even
more remarkable and it would be interesting to see if this pattern
continues for larger values of $n.$

In the Coulomb problem there are a number of local minima with energies
which are only very slightly above that of the global minimum, but have
completely different symmetries. In fact the number of stable local minima
appears to grow exponentially fast with $n$ \cite{EH}. The smallest number
of points for which such a metastable state occurs in the Coulomb problem
is $n=16$ and this is again mirrored in our findings, with a local minimum
produced at $n=16$ with $D_{4d}$ symmetry and an energy of $E=-24.9477$,
in comparison with the $T$ symmetric global minimum with energy $E=-24.9525.$
The existence of metastable states is one of the motivations for our use
of a simulated annealing algorithm combined with a number of different
random starting configurations.

\section{Comparison with a 3-particle interaction}\news
\label{sec-3pt}

As we saw in the previous section, if the points are constrained to lie
on the surface of a sphere then a good approximation
to the configurations which minimize our multi-particle energy are
obtained by minimizing the 2-particle Coulomb energy. The question
we address in this section is whether there exists a
2-particle or 3-particle energy which provides the same kind of
good approximation, in the sense of generating almost the same minimizing
configurations, but which has the additional features of being scale
invariant (in common with our multi-particle energy) and does not require
the points to be constrained to the surface of a sphere.

Clearly, there can be no scale invariant 2-particle energy, which
preserves rotational invariance, since the only possible quantity
from which to form an interaction is the distance between the two points.
We therefore turn our attention to a possible approximation involving a 3-particle 
interaction. 

In view of conjecture 3 applied inductively we can express our $n$-particle energy
function $E$ as a sum of \lq pure\rq\ $k$-particle energies $F_k$, for
$k=n,n-1,\ldots,3.$ For example, the conjectured inequality (\ref{enrec}),
leads us to define $F_n$ as
\be
F_n=E-\frac{1}{n-2}\sum E_i
\label{m71}
\ee
(and so $F_n\le 0$). Next we would define
\be
F_{n-1}=\frac{1}{n-2}\sum E_i-\frac{2}{(n-2)(n-3)}\sum E_{ij}
\ee
(where the second sum is over pairs $i>j$)
with $F_{n-1}\le 0.$
Continuing in this way we can express
\be
E=F_n+F_{n-1}+\ldots+F_3
\label{m72}
\ee
as a sum of (conjecturally) negative terms : the parts of the energy
which are not consequences of energies of proper subsets.

In usual physical models a complicated multi-particle energy function
can sometimes be expanded as a sum of pure $k$-point energies, starting with
$k=2$, and this 2-point energy may be the dominant factor. Here our expansion
(\ref{m72}) starts with $k=3$, but by analogy we might conjecture that this is
in some sense the \lq dominant part\rq. Thus we are led, very tentatively, to
compare our energy function $E$ with the 3-point energy function $F_3.$
In particular we will compare the minimal configurations of these two
energy functions. We will find remarkable agreement which will be quantified.

Although we have the (conjectured) inequality
\be
E\le F_3
\label{m73}
\ee
it will, for some purposes, be more enlightening to rescale $F_3$ by a
numerical constant. The reason is that
\be
F_3=\frac{1}{n-2}\sum E_I
\label{m74}
\ee
where $I$ runs over all (unordered) triples in $(1,2,\ldots,n)$ and
$E_I$ is the energy of the corresponding triangle of points in $\bR^3.$
But the number of such triangles (or triples) is just
\be
\bigg({{n\atop 3}}\bigg)=\frac{n(n-1)(n-2)}{6}
\ee
and so the {\em average energy per triangle} is
\be
E_\Delta=\frac{6}{n(n-1)(n-2)}\sum E_I=\frac{6}{n(n-1)}F_3.
\ee
Note that, as an approximation to $E$, $E_\Delta$ is very much
worse than $F_3.$ On the other hand it is a natural quantity to
measure numerically and in particular it will have a sensible
asymptotic behaviour for large $n,$ as we shall postulate.
Explicitly, $E_\Delta$ is given by the formula
\be
E_\Delta=\frac{-6}{n(n-1)(n-2)}\sum_{i>j>k}\log\{
\frac{1}{2}[
\cos^2(\theta_1/2)+
\cos^2(\theta_2/2)+
\cos^2(\theta_3/2)
]\} 
\label{edelta}
\ee
where $\theta_1,\theta_2,\theta_2$ are the angles in the triangle formed
by the three points $\bx_i,\bx_j,\bx_k,$ and we have used the explicit
formula (\ref{vtri}) for the 3-particle volume.

\begin{figure}[ht]
\begin{center}
\leavevmode
\ \hskip -2cm
\epsfxsize=14cm\epsffile{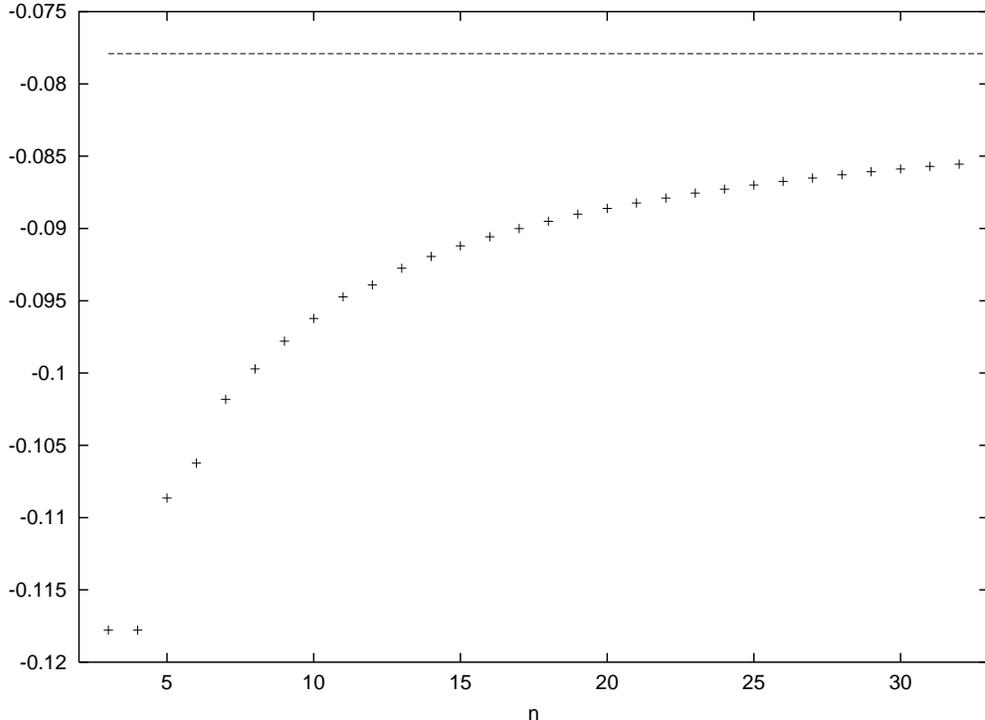}
\vskip -0.5cm
\caption{
The minimal value (crosses) of the energy $E_\Delta$ for $n$ points
in the range $3\le n\le 32.$ The dashed line represents the average value
of $E_\Delta$ for three randomly distributed points.
}
\label{fig-3pt}
\end{center}
\end{figure}

\noindent In fig.~\ref{fig-3pt} we plot (crosses) the minimal value of $E_\Delta$ for
$n$ points with $3\le n\le 32.$ 
The precise values are presented in the fifth column of table~1.
Again these results are obtained using a
multi-start simulated annealing code. For $n=3$ and $n=4$ the minimal value
of $E_\Delta$ is obviously $-\log(9/8),$ since in these two cases all triples
of points can simultaneously be made to form equilateral triangles;
this being achieved by the regular tetrahedron in the case of four points.
For $n>4$ the energy rises, though rather slowly, and already at $32$ points
it has clearly begun to level off. 
It seems probable that an
 upper bound for the energy $E_\Delta$
is obtained by the average value of the 3-particle energy for three
randomly distributed points. By numerically computing the values for 
50,000 randomly chosen triangles we find that this average value is
$\bar E_\Delta=-0.078.$ The constant $\bar E_\Delta$ is indicated in fig.~\ref{fig-3pt}
as a dashed line, and is consistent with the numerical results. 
Further simulations, for larger values of $n$ are required to determine whether
the minimal value of $E_\Delta$ asymptotes to the value $\bar E_\Delta,$ given
by randomly distributed points, or whether it tends to a slightly lower level.

The aim of this section was to find an energy which was a sum of 3-particle
interactions such that the arrangement of minimizing points closely reproduces
the earlier results based on the multi-particle energy $E$. 
The energy $E_\Delta$ succeeds in this aim, with the positions of all points of the 
$E_\Delta$ minimizing configurations being within $1\%$ of those in the $E$
minimizing arrangements. A more quantitative comparison is made is table~1
where we list the quantity $\widetilde E,$ which is the value calculated for 
the energy $E$ from the $E_\Delta$ minimizing configuration. A comparison 
between $E$ and $\widetilde E$ demonstrates the close match between the
two sets of configurations and confirms that $E_\Delta$ is a good approximation.
The worst match is for $n=5,$ where the three points which form an equilateral
triangle lie on a circle whose radius is increased by less than $\frac{1}{2}\%$
in comparison with the $E$ minimizing configuration, though otherwise these two
configurations are identical. In the last column of table~1 we list,
for comparison with $\delta,$
the deformation from a spherical arrangment, $\widetilde\delta$, (computed
in the same way as $\delta$) for the $E_\Delta$ minimizing points.

The energy function $E_\Delta$ helps us to understand why the configurations
we produce are spherical polyhedra with faces which are generally triangular.
This is because, locally, the arrangment of points favours equilateral triangles,
and if we consider introducing an additional point, far from a current spherical
distribution, then clearly all the triangles involving this point will have 
one very acute angle, which can be increased by moving the point closer to the
shell, hence lowering the energy and producing an attraction.

It is intriguing that our 3-particle volume (\ref{vtri}) has arisen
earlier in a different physical context \cite{Cal}. This concerns quantum
many-body problems for which eigenstates can be found explicitly. It turns
out that the addition of precisely this 3-body interaction allows the
explicit computation of some eigenstates (including the ground state)
of less tractable Hamiltonians with only 2-body forces. In the quantum
system the Hamiltonian involves a summation of the 3-particle volume over
 all triples of particles, whereas in our application the energy involves
taking the product of the 3-particle volumes over all triples, and then taking
minus the logarithm. However, the difference between a sum and product is not as
 radical as it might at first appear, since the interaction produced by the 3-particle volume
is very weak. In fact we have investigated replacing the product of volumes 
by a sum and found that the results are qualitatively robust to this
modification. For this modification there is a bound derived in \cite{Cal}
for the 3-particle energy per triangle in the limit $n\rightarrow\infty$ and this
agrees quite well with our expected upper bound
$\bar E_\Delta=-0.078.$ Explicitly, the bound of \cite{Cal}, which applies to the
modified energy, is $-\log(\frac{3}{4}+\frac{1}{\pi})=-0.066.$
It would be interesting to investigate this connection further, and
to determine whether our multi-particle volume also leads to tractable
Hamiltonian systems.

\section{The unconstrained problem}\news
\label{sec-prods}

In Section 3 we defined a complex-valued function $D$ on the open set
$t_{ij}\ne t_{ji}$ in a product of $n(n-1)$ 2-spheres. We then restricted
it to half the number of 2-spheres by taking $t_{ji}$ to be the anti-pode
of $t_{ij}.$ The energy function $E$ on ${\cal C}_n(\bR^3)$ which we have
been studying was defined by
\be
E=-\log|D|
\ee
evaluated on the $t_{ij}$ defined by $(\bx_1,\ldots,\bx_n)$
as in (\ref{unitv}). We have the conjectured inequality
$E\le 0$ and we have, in previous sections, been investigating
the configurations $(\bx_1,\ldots,\bx_n)$ which minimize $E.$
We now ask how do these compare with the configurations of $t_{ij}$
which minimize $E,$ without the constraint required by (\ref{unitv}),
that the $t_{ij}$ originate from a configuration of points in $\bR^3.$
Note that the space of $t_{ij}$ has real dimension $n(n-1)$, while
${\cal C}_n(\bR^3)$ has dimension $3n$ (and reduces to $3n-4$ when
we factor out by translation and scale, which do not affect $E$).
Thus there are many constraints. 
For $n=3$ we have
\be
n(n-1)=6, \quad 3n-4=5
\ee
so that there is one constraint. It is that the 3 points
$t_{12},t_{13},t_{23}$ lie on a great circle
(and also form an acute-angled triangle). It therefore seems to be
quite remarkable, and presumably significant, that a numerical
investigation suggests that the minimum energy of the constrained
and unconstrained problems coincide, as do the corresponding
configurations. This may well be a clue to explaining why our energy
minimizing configurations have such striking properties.

Although the constrained and unconstrained problems appear to have the
same minima for the energy, the corresponding statement for the maxima
is definitely false. In fact, for the unconstrained problem the energy
can be infinite or equivalently the function $D$ can vanish. This happens
already for $n=3$ with $t_{12}=t_{23}=\infty$ and $t_{13}=0$, so that
the 3 polynomials $p_i$ all coincide (having $0,\infty$ as the 2 roots)
and hence are linearly dependent.

One might be tempted to combine the unconstrained problem of this section with
the 3-point energy of the previous section, and determine the unconstrained
configurations of $t_{ij}$ which minimize the average energy per triangle
$E_\Delta,$ summed over all triples of $t_{ij}$ with $i<j.$ However, a numerical
study reveals that for $n>3$ the minimal configurations for this problem do not resemble
our earlier configurations.

\section{The planar restriction}\news
\label{sec-plane}

It is interesting to investigate the minimal energy configurations
for points in the plane, that is, we restrict our configuration
space to 
$(\bx_1,\ldots,\bx_n)\in{\cal C}_n(\bR^2).$
Again we can make use of the translation and scale invariance of our problem
to fix the centre of mass at the origin and make all the points lie
in the unit disc $|\bx_i|\le 1,$ with at least one point on the boundary.

\begin{figure}[ht]
\begin{center}
\leavevmode
\ \hskip -2cm
\epsfxsize=12cm\epsffile{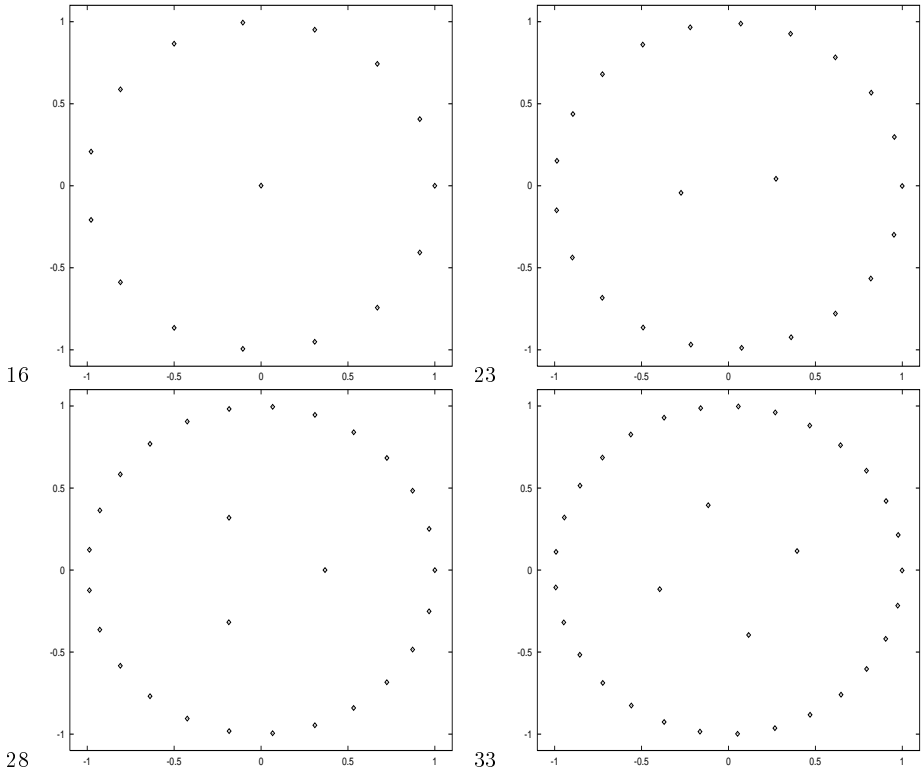}
\vskip -5cm
\caption{
The positions of the energy minimizing points in the plane
for the jumping values $n=16,23,28,33.$
}
\label{fig-2d}
\end{center}
\end{figure}

The results for $2\le n\le 15$ are perhaps not surprising, with the
$n$ points lying on the unit circle and forming the vertices of a regular 
$n$-gon. However, for $n=16$ the minimal energy configuration consists
of a regular 15-gon on the unit circle and a single point at the origin.
These points are displayed in fig.~\ref{fig-2d}.16.

The pattern of an $(n-1)$-gon plus one point at the centre continues
until $n=23,$ at which point the configuration comprizes a 21-gon
plus two points placed in the interior of the disc equidistant from the origin
 and lying on a diameter. This is displayed in fig.~\ref{fig-2d}.23.

At $n=28$ a further point enters the interior of the disc, 
producing a 25-gon with an equilateral triangle inside;
fig.~\ref{fig-2d}.28.
At $n=33$ another point enters the interior of the disc, leading to a 29-gon 
with a square arrangement at the interior; fig.~\ref{fig-2d}.33.
 Note that the points in the interior
are always arranged in the minimal energy configuration of that
number. 

We refer to the sequence of numbers, $n=16,23,28,33,\ldots$\
at which an additional point enters the interior of the disc,
as the jumping values.

In addition to the global minimum energy configurations 
described above we have found a large number of local
minima, associated with one or more points moving
from the interior of the disc to the boundary, or vice-versa.
A large number of annealing runs were made in each case to
find all the local minima and hence determine the global ones.

Our energy function is scale invariant, so it is interesting
that for $n>22$ a scale  has emerged, given by the 
ratio of the radius
of the interior circle to that of the bounding disc. Presumably
these radii, which increase slightly with $n,$ have some universal
geometrical character.

It is useful to repeat our earlier comparisons with 2-particle
and 3-particle interactions, applied to the planar case.

First, we consider the Coulomb interaction. If $n$ point-particles in 
the unit 3-ball interact via the 2-particle Coulomb energy (\ref{coul})
then the minimal energy configurations consists of points which all lie
on the unit sphere. The relevant theorem here is that a harmonic function
on a bounded domain takes its minimal value on the boundary of the domain.
However, if we consider the planar version, so that the points are
confined to the unit disc, then the configuration of points which
minimizes the Coulomb energy (\ref{coul}) can include points which
lie in the interior of the disc. This is because the Coulomb energy
is a harmonic function in three-dimensions but not in two-dimensions.
If the Coulomb energy is replaced by a logarithmic energy then this
is harmonic in two-dimensions and so all points lie on the unit
circle for the global minimizer.

The configurations which are the global minima for the Coulomb energy 
of points in a disk have been computed numerically for up to 80 points
in \cite{Nur}. A glance at the figures in this reference show an
immediate qualitative similarity with those in fig.~\ref{fig-2d}. For
$n<12$ the points form a regular $n$-gon on the unit circle. For
$12\le n\le 16$ they form an $(n-1)$-gon with a single point at the centre
of the disc. For $n=17,18$ there are two points in the interior of the
disc which lie on a diameter, in the same manner as in fig.~\ref{fig-2d}.23.
For $n=19,20,21$ there are three interior points which form an
equilateral triangle,  as in fig.~\ref{fig-2d}.28. For $n=22,23,24$
there are four interior points on the vertices of a square, which
is the arrangment in fig.~\ref{fig-2d}.33. Thus it seems that the same
patterns emerge in the Coulomb case as for our energy minimizers,
though the jumping values, at which additional points enter the
interior of the disc, are shifted to the sequence $n=12,17,19,22,\ldots$
Thus the jumps occur earlier and more frequently for the Coulomb energy.
For larger values of $n$ the number of shells (circles on which
$m$ points approximately lie on a regular $m$-gon) increases beyond
two, so that, for example, at $n=80$ there are four shells containing
4,10,18,48 points respectively, working from the inner shell out to
the boundary of the disc. We therefore expect that 
for larger $n$ a similar pattern
of increasing shells will emerge as the minimizers of our geometric
energy, though the number of points required to generate a given number
of shells will be larger than for the Coulomb energy.

The above results suggest that the planar case is more sensitive to the
form of the energy function than the three-dimensional problem, since 
in three-dimensions we found a much closer agreement between the minimizers
of our energy and those of the Coulomb energy. This may be due to the fact
that points interacting via a Coulomb energy automatically lie on the
surface of a sphere, so there is no issue of matching the number of interior
points. This explanation appears likely, given that the arrangements
of interior and boundary points also agrees for the two energies
 in the planar case, it being the {\em number} of interior points which fails
to match.

We can make use of the sensitive nature of the planar restriction to
investigate whether there is a power $p,$ in the more general 2-particle
energy (\ref{gen}) for which a better approximation to our multi-particle
energy configurations can be achieved than the Coulomb case of $p=1.$
This indeed turns out to be the case. First of all, requiring the first
jumping value to be at $n=16$ determines that $p$ lies in the small range
$0.58<p<0.64.$ If $p$ lies outside this range then the jump to a single
point in the centre of the disc occurs at a smaller or larger value of $n$
than 16. Although the second jumping value, $n=23,$ can also be matched, we
find that there is no value of $p$ within this range
 such that the third jumping value occurs
at $n=28.$ Thus, there is no value of $p$ such that the energy minimizing
configurations agree for all $n,$ but choosing $p=0.6$ gives a good approximation,
with the jumping values being $n=16,23,27,31,\ldots.$ The arrangements
of points for values of $n$ which have the same number of interior points
are essentially identical.

An obvious question is to ask about the minimal arrangements of points
on a sphere interacting with the energy (\ref{gen}) for the selected 
value of $p=0.6.$ Computing these configurations we find that they agree with
those of the Coulomb energy, $p=1,$ to within the accuracy that we made the
earlier comparison with our energy function. Given our earlier comments, then 
for much larger values
of $n$ it might be expected that a noticeable difference may emerge between
the case $p=0.6$ and $p=1.$ As far as we are aware the minimal energy
configurations for the more general energy (\ref{gen}) have been studied 
only for $p\ge 1.$

If we compare with the three-particle energy, $E_\Delta$ 
given by (\ref{edelta}), restricted to the planar case we again find the same
patterns of points, but the precise jumping values are even further
away from those of the multi-particle energy than for the Coulomb
approximation. Explicitly, the jumping values are $n=8,12,14,16\ldots.$

\section{Points in hyperbolic space}\news
\label{sec-hyp}

If we replace Euclidean space $\bR^3$ by hyperbolic 3-space, which we
denote by $\HH^3_\kappa$, where $-\kappa^2$ is the curvature, then 
there is an analogue of the Berry-Robbins problem, together with 
all the other issues we have addressed in the Euclidean case,
such as the configurations of points which minimize
a geometrical multi-particle energy.
This was already noted in \cite{At1,At2}.

The natural generalization to hyperbolic space is to ask for
a map
\be
F_n : {\cal C}_n(\HH^3_\kappa)\mapsto GL(n,\bC)/(\bC^*)^n\,.
\label{defmaph}
\ee
which is compatible with the action of $\Sigma_n$
and $SL(2,\bC),$ where this acts 
(modulo $\pm 1$) on hyperbolic space
as its group of (oriented) isometries and on $GL(n,\bC)$ via
the irreducible $n$-dimensional representation.

\noindent The construction of Section \ref{sec-map}, based on
the polynomials $p_1,\ldots,p_n$ can be repeated  in 
hyperbolic space in a similar
way. Given two distinct points $\bx_i,\bx_j\in\HH^3_\kappa,$
we define $t_{ij}$ to be the point on the Riemann
sphere at  infinity
along the oriented geodesic through $\bx_i$ and $\bx_j.$
Note that in the hyperbolic case $t_{ij}$ and $t_{ji}$
are no longer antipodal points on the Riemann sphere:
in fact the notion of anti-pode requres us to fix an origin in
$\HH^3_\kappa,$ and is not $SL(2,\bC)$ invariant.

We take the projective model of $\HH^3_\kappa,$
as the interior of the 3-ball of radius $1/\kappa$
in $\bR^3.$ The geodesics are then just straight lines,
and $t_{ij}$ is  the point where the (oriented)
line $\bx_i\bx_j$ meets the sphere of radius $1/\kappa$
in $\bR^3.$ In the zero curvature limit, $\kappa\to 0,$
of the hyperbolic construction, we recover the earlier
Euclidean case.

A geometric proof of the independence of the polynomials
$p_1,\ldots,p_n$ can be given for $n=3$ \cite{At1},
but the hyperbolic version of conjecture 1 still remains
unproven for $n>3.$ 

The complex function $D$ of the $t_{ij}$ introduced in Section 3
is still well-defined because we always have $t_{ij}\ne t_{ji}.$
It is covariant under the full isometry group of hyperbolic space
with reflections inducing complex conjugation.
Assuming $D\ne 0$ (the hyperbolic conjecture) we can again define
a volume $V=|D|$ and an energy function
\be
E=-\log V.
\ee
This definition is $SL(2,\bC)$ invariant and so is intrinsic
to hyperbolic space.

The numerical evidence again supports the hyperbolic version of
conjecture 2, namely $V\ge 1$ or $E\le 0.$

Turning now to the minimizers of the energy $E=-\log V,$
we find that the same arrangements of points
as in Euclidean space are the minimizers, but there is now an intrinsic
scale, provided by the curvature of hyperbolic space, and the energy is
minimized in a singular limit as the overall scale of the configuration
tends to zero. In other words the curvature of hyperbolic space provides
an attractive force.
This can be illustrated by considering the simple case
of an equilateral triangle with a varying scale.

\begin{figure}[ht]
\begin{center}
\leavevmode
\epsfxsize=13cm\epsffile{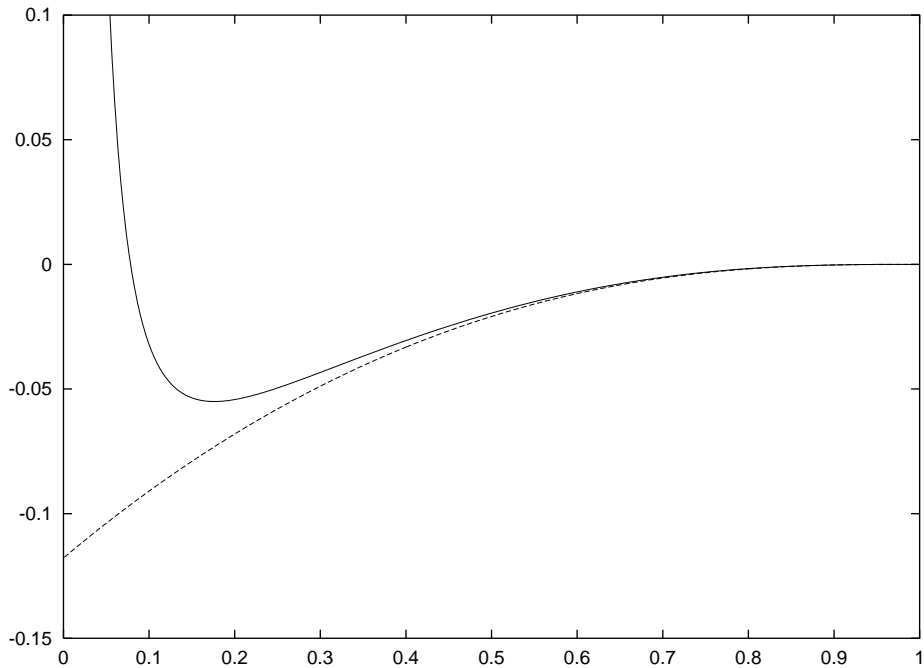}
\caption{Two quantities (as a function of $\rho$)
 associated with three points on the vertices of an 
equilateral triangle of scale $\rho/\kappa$ in hyperbolic space. The 
dashed curve is the energy $E$ of the static configuration. The solid curve is
the total energy $U$ for a rotating triangle with angular momentum $l=0.02.$ }
\label{fig-hyp}
\end{center}
\end{figure}

Consider the equilateral triangle with vertices on the
circle centered at the origin and of radius $\rho/\kappa,$ where $0<\rho<1.$
An explicit calculation of the energy for this configuration yields
\be
E=-\log\bigg\{\frac{3\sqrt{3}(\sqrt{12-3\rho^2}-\rho)}
{2(4-\rho^2)^{3/2}}\bigg\}\,.
\label{ehyptri}
\ee
This energy, as a function of $\rho,$ is displayed in fig.~\ref{fig-hyp}
as the dashed curve.
The minimum is attained in the singular limit $\rho\rightarrow 0,$
where the Euclidean result $E=-\log(9/8)$ is recovered. 
The scale invariance of the Euclidean case emerges by taking the limit
$\kappa\rightarrow 0$ and $\rho\rightarrow 0$ such that the ratio
$\rho/\kappa$ is finite, and gives the arbitrary scale of the triangle.

It may be possible, by the addition of an extra repulsive
contribution to the energy, to balance the attraction induced
by hyperbolic space, but, as yet, we have not found
an elegant way to incorporate such a modification.

An alternative to the addition of extra terms is to exploit the
attraction of hyperbolic space to construct bound orbits of 
rotating configurations. As an example, let us reconsider the
above configuration of three points on an equilateral triangle 
with scale $\rho\in(0,1),$ where for simplicity we set $\kappa=1.$
Keeping the plane of the triangle fixed, introduce an angle $\phi$
describing the freedom to rotate the triangle about its centre.
3-particle configurations have a fixed point set under the dihedral group
$D_{3h}$ which is two-dimensional, and is parametrized by the coordinates
$\rho$ and $\phi$ which we have introduced. When discussing time dependent
solutions it is therefore a consistent reduction to restrict to the two-dimensional
dynamical system given by $\rho(t),\phi(t).$ Taking the point particles to have unit mass,
the Lagrangian for this system is given by
\be
{\cal L}=\frac{3}{2(1-\rho^2)^2}(\dot\rho^2+(1-\rho^2)\rho^2\dot\phi^2)-E
\ee
where $E$ is the potential energy given by (\ref{ehyptri}) and dot denotes
differentiation with respect to the time coordinate $t.$ 

The expression for the conserved angular momentum is
\be
l=\frac{\rho^2\dot\phi}{(1-\rho^2)}.
\ee
For a solution to exist that describes a triangle with fixed radius which rotates
at constant angular velocity $\dot\phi$ therefore requires a solution of the equation 
\be
\frac{d U}{d \rho}=0, 
\quad \mbox{where}\quad
U(\rho)=E+\frac{3l^2(1-\rho^2)}{2\rho^2}.
\ee
Providing the angular momentum $l$ is not too large, then solutions of this
equation indeed exist with $\rho\in(0,1).$ 
For example, for $l=0.02$ the total energy
$U(\rho)$ is plotted (solid curve) as a function of $\rho$ in fig.~\ref{fig-hyp}.
This function has a clear minimum (which occurs at $\rho=0.176$) demonstrating the
existence of a dynamical solution describing three orbiting points on the vertices
of an equilateral triangle associated with this finite non-zero scale. The fact that
this critical point is the global minimum also shows that this orbit is stable within
the class of rotating triangular solutions. Similarly, one expects more complicated
orbiting configurations to exist for larger numbers of particles.

In Euclidean space reflection in an origin interchanges $t_{ij}$ and $t_{ji}$,
since these are anti-podes, and the function $D$ gets conjugated by such a reflection.
Thus interchanging the roles of the two indices in $t_{ij}$ takes $D$ into $\bar D.$
This is not a trivial observation since we get the $p_i$ by symmetrizing over $j$
and then (taking the exterior product) we skew-symmetrize over $i.$ In hyperbolic
space, where we have no usual notion of anti-pode, we cannot argue in this way. In fact
reversing the roles of the indices in $t_{ij}$ produces a quite {\em different
function} $D^\dagger.$

As noted at the end of \cite{At2} hyperbolic geometry is closely related to Minkowski
geometry because of the isomorphism (of connected groups)
\be
SL(2,\bC)\cong \mbox{Spin}(3,1).
\label{m103}
\ee
In \cite{At2} it was shown how to attach points $t_{ij}$ to $n$ points (or events)
in Minkowski space {\em with their past histories} (or world-lines).
Although not entirely obvious it is true (and was observed in \cite{At2}) that
$t_{ij}\ne t_{ji}.$ Hence, proceeding as in Section 3, we can still define our
complex-valued function $D$ (which may now have zeros). 
It has full Lorentz-invariance because of (\ref{m103}), but we have to be careful 
about the disconnected components corresponding to space reflection and time-reversal.
As before, space-reflection takes $D$ into $\bar D$, but time-reversal is, in general,
not allowable since it would take past histories into the (unknown) future.
However, we can apply it in the very special case of particles travelling along
straight-lines (ie. uniform motion) and all emerging from a \lq big-bang\rq\ in the past.
As shown in \cite{At2}  this essentially reduces to pure hyperbolic geometry and 
time-reversal in this case makes sense (with a forecast future of continuing uniform
motion) and takes $D$ into the different function $D^\dagger$.
In fact this holds provided each pair of world-lines are (and remain) coplanar - 
the motion need not be uniform.

This complex function $D,$ depending on past histories, raises interesting
physical questions and its implications will be considered more carefully on a future
occasion.
It has been suggested to us by Michael Berry that it may be related to
the use of retarded potentials in the relativistic treatment of the
dynamics of charged particles.

\section{The complex phase}\news
\label{sec-real}

\begin{figure}[ht]
\begin{center}
\leavevmode
\epsfxsize=13cm\epsffile{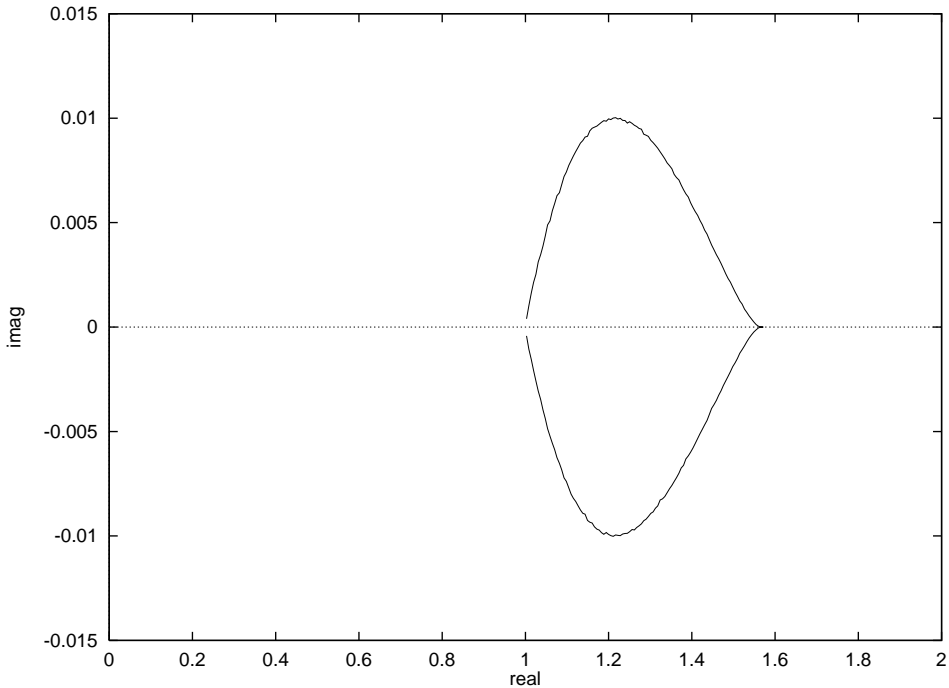}
\caption{The curves bounding the region of the complex plane where
 ${D}$ takes its values for $n=4.$}
\label{fig-cp4}
\end{center}
\end{figure}

So far in this paper we have made extensive use of the real volume,
$V=|D|,$ but we have ignored the phase of the complex function $D,$
associated with the volume form.

As explained in Section 3, if a configuration of $n$ points 
in $\bR^3$ has a reflection symmetry
then $D$ is automatically real,
and so $D$ is always 
real for $n<4.$ It is to be expected that for a general arrangment of four
or more points then  $D$ will be complex, but the region of the
complex plane where $D$ takes its values is by no means obvious.
In this brief section we investigate this aspect for the simplest case of
four points. By computing $10^7$ random configurations of four points
we have been able to map out the region of the complex plane in which
${D}$ is constrained to lie for $n=4.$ It is the interior of the
region bounded by the solid curves in fig.~\ref{fig-cp4}.
Note the differing scales on the real and imaginary axes, indicating that
the complex part of $D$ is very small for all configurations
of four points; in fact the phase angle is always less than one degree, which
is quite surprising. An obvious question concerns the configuration of points
for which the phase angle is maximal, and this has an amusing answer.
Recall that the phase is zero if there is a reflection symmetry, so in some
sense this configuration should be maximally asymmetric, but in a controlled
rather than random way. This is precisely what happens, with the four points
arranged in two groups of two and lying in two parallel planes. The angle
between the lines joining each of the coplanar points is $45^\circ$ and it
is in this sense that the asymmetry is maximal, since if this angle is
zero then all four points are coplanar and hence the phase is zero. Similarly
if this angle is $90^\circ$ then again there is a reflection symmetry, this time
the reflection plane contains two of the coplanar points and the reflection
leaves these two points fixed and exchanges the remaining two. Thus the
observed $45^\circ$
angle is precisely mid-way between two possible values of the angle
 for which there is a reflection symmetry.

We have performed similar calculations for more points and the results
are qualitatively similar, with the allowed region of the complex plane
having a shape of the same form as for $n=4,$ but being of slightly larger
area. Consequently the maximal value of the phase grows slightly with $n,$
and more complicated maximally asymmetric configurations occur.

\section{Conclusions}\news

This paper began with the aim of numerically verifying conjecture 1, establishing
the existence of a certain natural map
${\cal C}_n(\bR^3)\mapsto U(n)/U(1)^n.$
For this purpose we introduced a complex function $D$ on the configuration space
${\cal C}_n(\bR^3)$ and the associated \lq energy\rq\ function
$E=-\log|D|.$ Numerical simulations applied to $|D|$ not only verified
(for $n\le 20$) the inequality $|D|>0$ (implying conjecture 1), but also suggested
the stronger inequality $|D|\ge 1,$ with equality only for collinear configurations.
This encouraged us to make our second conjecture, which in terms of the energy asserts
that $E\le 0.$

Calculations for $n=3$ then suggest a more precise inductive inequality
\be
E\le \frac{1}{n-2}\sum E_i
\label{m121}
\ee
where $E_i$ is the energy of the configuration of $n-1$ points obtained
by omitting $\bx_i.$ Numerical calculations substantiated this and so
encouraged us to make (\ref{m121}) our third conjecture.

We were then led, out of pure curiosity, to ask the opposite question
about the minimum value of $E$ and the configurations which produce this
minimum. Very surprisingly, we found them (up to $n=32$) to be approximately
spherical polyhedral structures of high symmetry. 
It appears that these
particular polyhedra (or their duals), comprizing
$n$ vertices and $2(n-2)$ triangular faces
forming $12$ pentamers and $n-12$ hexamers, are somewhat generic configurations.
They arise in a number of complicated 3-dimensional physical applications
(eg. Skyrmions, Fullerenes), as well as in the 2-dimensional problem for
configurations of Coulomb charges on a sphere.

Given the basic geometric nature of our problem, with its energy function,
it may well be that it is the simplest example of a whole universality
class of similar effective interactions. It may therefore be a useful
model and can perhaps act as a simple approximation to more complicated
physical systems. For example it should be possible to use our minimum
energy configurations as a basis for predicting the structure of
higher charge Skyrmions and even of constructing them.

In an attempt to understand why our energy function leads to these kind of
minima we were led to look at two simplifications of the problem.
In the first we replaced our energy function by its associated 3-point energy,
by taking the energy function for $n=3$ (given by a simple explicit formula)
and summing it over all triples in our set of $n$ points.
We found (numerically) essentially the same minimum energy configurations,
indicating that this 3-point energy is in some sense the dominating part of
the total energy.

The second simplification we made was to compute the energy function for 
all sets of points $t_{ij}$ on the 2-sphere (with $t_{ji}$ the anti-pode
of $t_{ij}$), without requiring the constraint that they arise as the 
set of directions joining points $\bx_i,\bx_j$ of a configuration in
${\cal C}_n(\bR^3).$ We refer to this as the unconstrained energy problem and
again (numerically) we found the same set of minima.

All these results seem to indicate some underlying and very stable
phenomenon, and following up on the various simplifications may yield
a better theoretical understanding of what is, at present, computer
evidence. There are a number of very interesting mathematical results
concerning the space of shapes of polyhedra and triangulations of the
sphere \cite{Thu} and these may prove useful in further theoretical
investigations. 

Finally we investigated two variants of our energy problem.
One is to restrict to planar configurations, and the other is to replace
Euclidean space by hyperbolic space. The latter may have interesting
consequences in Minkowski space.\\

{\sl Note Added.}

Recently Eastwood and Norbury \cite{EN} have proved conjecture 1
for the case $n=4.$ Their method involves using MAPLE to generate
an expression involving several hundred terms which they then neatly
rearrange into geometrical objects, such as the volume of
the tetrahedron formed by the four points. The inequality that
they derive is also very close to proving conjecture 2 for $n=4.$\\

\section*{Acknowledgements}\news
Many thanks to Richard Battye for useful discussions.

\noindent PMS acknowledges the EPSRC for an advanced fellowship.


\begin{thebibliography}{99}

\bibitem{At1}
M.F. Atiyah, 
{\em The Geometry of Classical Particles},
Surveys in Differential Geometry (International Press) 7, 1 (2001).

\bibitem{At2}
M.F. Atiyah, 
{\em Configurations of Points},
Phil. Trans. R. Soc. Lond. A 359, 1 (2001).

\bibitem{BS} R.A. Battye and P.M Sutcliffe,
{\em Symmetric Skyrmions}, 
Phys. Rev. Lett. 79, 363 (1997);\
{\em Solitonic Fullerene Structures in Light Atomic Nuclei},
Phys. Rev. Lett. 86, 3989 (2001);\
{\em Skyrmions, Fullerenes and Rational Maps}, hep-th/0103026.

\bibitem{BSTK}
B. Berger, P.W. Shor, L. Tucker-Kellogg and J. King,
{\em Local Rule-Based Theory of Virus Shell Assembly},
Proc. Natl. Acad. Sci, 91, 7732 (1994).

\bibitem{BR}
M.V. Berry and J.M. Robbins,
{\em Indistinguishability for Quantum Particles:
Spin, Statistics and the Geometric Phase},
Proc. R. Soc. A 453, 1771 (1997).

\bibitem{Cal}
F. Calogero and C. Marchioro,
{\em Exact bound states of some $N$-body systems},
J. Math. Phys. 14, 182 (1973).

\bibitem{EN}
M. Eastwood and P. Norbury,
{\em A proof of Atiyah's conjecture on configurations
of four points in Euclidean three-space},
preprint (2001).

\bibitem{JRE}
J.R. Edmundson, 
{\em The Distribution of Point Charges on the Surface of a Sphere},
Acta Cryst. A48, 60 (1992).

\bibitem{EH}
T. Erber and G.M. Hockney,
{\em Complex Systems: Equilibrium Configurations of $N$
Equal Charges on a Sphere},
Adv. Chem. Phys. 98, 495 (1997).

\bibitem{KHOC}
H.W. Kroto, J.R. Heath, S.C. O'Brien, R.F. Curl and R.E. Smalley,
{\em $C_{60}$ - Buckminsterfullerene},
Nature 318, 162 (1985).

\bibitem{MKS}
T.W Melnyk, O. Knop and W.R. Smith,
{\em Extremal Arrangements of Points and Unit Charges on a Sphere:
Equilibrium Configurations Revisited},
Can. J. Chem. 55, 1745 (1977).

\bibitem{Nur}
K.J. Nurmela, 
{\em Minimum-Energy Point Charge Configurations on a Circular Disk},
J. Phys. A 31, 1035 (1998).

\bibitem{JJT}
J.J. Thomson, 
{\em On the Structure of the Atom},
Philos. Mag. 7, 237 (1904);\
{\em On the Structure of the Molecule and Chemical Combination},
Philos. Mag. 41, 510 (1921).

\bibitem{Thu}
W.P. Thurston,
{\em Shapes of Polyhedra and Triangulations of the Sphere},
Geometry and Topology Monographs 1, 511 (1998).

\bibitem{sabook} 
P.J.M. van Laarhoven and E.H.L. Aarts, {\em
Simulated Annealing: Theory and Applications}, Kluwer Academic Publishers, (1987).


\end{thebibliography}
\end{document}